\documentclass[onecolumn,superscriptaddress,nofootinbib,preprintnumbers]{revtex4-2}
\usepackage[colorlinks=true,breaklinks=true]{hyperref}
\usepackage[utf8]{inputenc}
\hypersetup{urlcolor=Blue,
	    citecolor=Blue,
	    linkcolor=Blue}
\usepackage{mathtools}
\usepackage[dvipsnames]{xcolor}
\usepackage{bbm}
\usepackage[normalem]{ulem}
\newcommand{\CNB}{C$\nu$B\;}

\newcommand{\SVcomment}[1]{\emph{\color{red}SV:#1}}

\newcommand{\CD}[1]{{\leavevmode\color{red}{#1}}}

\begin{document}

\preprint{ULB-TH/23-05}

\title{Testing secret interaction
with astrophysical neutrino point sources}

\author{Christian D\"oring}
\email{christian.doring@ulb.be}
\affiliation{Service de Physique Theorique, Universite Libre de Bruxelles, Boulevard du Triomphe, CP225, 1050 Brussels, Belgium}

\author{Stefan Vogl}
\email{stefan.vogl@physik.uni-freiburg.de}
\affiliation{Institute of Physics, University of Freiburg,\\Herrmann-Herder-Str.~3, 79104 Freiburg, Germany}

\date{\today}


\begin{abstract}
\textbf{Abstract:} 
Recently, the IceCube collaboration observed a neutrino excess in the direction of NGC 1068 with high statistical significance.
This constitutes the second detection of an astrophysical neutrino point source after the discovery of a variable emission originating from the blazar TXS~0506+056. Neutrinos emitted by these sources traverse huge, well-determined distances on their way to Earth. This makes them a promising tool to test new physics in the neutrino sector. We consider secret interactions with the cosmic neutrino background and discuss their impact on the flux of neutrino point sources. The observation of emission from NGC 1068 and TXS 0506+056 can then be used to put
limits on the strength of the interaction. We find that our ignorance of the absolute neutrino masses has a strong impact and, therefore, we present limits in two benchmark scenarios with the sum of the neutrino masses around their lower and upper limits.
\end{abstract}

\maketitle


\section{Introduction\label{sec:Intro}}

Our Universe is home to billions of galaxies some of which host powerful cosmic particle accelerators. These objects are the sources of high energy cosmic rays, e.g. protons, nuclei, gamma-rays and neutrinos. High energy protons and nuclei with energies up to $10^{10}$ GeV have been observed for a long time \cite{Letessier-Selvon:2011sak}. However, they are affected by magnetic fields on their way to Earth and thus we cannot trace them back to the objects from which they originate. Gamma-rays with energy above $\mathcal{O}(10)$ TeV are absorbed over cosmological distances  \cite{DeAngelis:2013jna} and can be dampened further by the local distribution of matter in and around the source. Therefore, neutrinos are sought-after messengers that are expected to allow further insights into the nature and distribution of the  sources of high energy cosmic rays.
Recently, the IceCube collaboration presented evidence for the emission of neutrinos from the active galaxy NGC 1068 \cite{IceCube:2022der}. This constitutes the second discovery of an astrophysical neutrino point source after the blazar TXS 0506+056 \cite{IceCube:2018cha,IceCube:2018dnn}. This places us right at the beginning of the era of neutrino astronomy.

In the future, the measurement of high energy neutrinos of astrophysical origin and the determination of their sources will improve and allow us to deepen our knowledge of both, the astrophysics of these objects and the particle physics involved. Potentially, in these systems physics beyond the Standard Model (SM) shows up, providing in this way the window to unexplored regions in the microscopic world.
A possibility we want to study here is a coupling to a light mediator, a scenario which is also known as secret neutrino interactions \cite{1991PhLB..267..504C,Acker:1990zj,Beacom:2004yd}.  Such an interaction is surprisingly hard to test in the lab \cite{Lessa:2007up,Laha:2013xua,Berryman:2018ogk,Heeck:2018nzc,Babu:2019iml,Blinov:2019gcj,Brdar:2020nbj,Lyu:2020lps,Berbig:2020wve}. Therefore, various effects on cosmology and astrophysical systems have been discussed in the literature in an attempt to constrain this scenario, see e.g. \cite{Nussinov:1982wu,Kolb:1987qy,Cyr-Racine:2013jua,Huang:2017egl,Kreisch:2019yzn,Barenboim:2019tux,Forastieri:2019cuf,Shalgar:2019rqe,RoyChoudhury:2020dmd,Escudero:2021rfi,Chang:2022aas}. 
Neutrinos from astrophysical sources provide one means to search for secret interactions. These neutrinos need to traverse large distances on their way to Earth. Within the SM this is not an issue since the probability of an interaction between neutrinos and matter is very small. However, in the presence of new, beyond the SM interactions this can change. 
Concretely, interactions of the astrophysical neutrinos with the cosmic neutrino background (\CNB) can turn the universe opaque to astrophysical neutrinos.  Thus the observation of astrophysical neutrinos can be used to constrain secret interactions. This idea has received some attention since IceCube first detected astrophysical neutrinos, see e.g. \cite{Ioka:2014kca,Ng:2014pca,DiFranzo:2015qea,Blum:2014ewa,Araki:2015mya,Kelly:2018tyg,Bustamante:2020mep,Esteban:2021tub,Murase:2019xqi,Carpio:2021jhu,Hooper:2023fqn}.
Similar effects due to the interaction with dark matter have been considered in \cite{Arguelles:2017atb,Pandey:2018wvh,Choi:2019ixb,Alvey:2019jzx,Ferrer:2022kei}.
In most studies the whole diffuse astro-neutrino flux is considered to include all the available data. Here, however, we want to follow the line of thought of Ref.~\cite{Kelly:2018tyg}  and study the neutrinos associated with individual sources. 
NGC 1068 data was recently found to constrain active-sterile mixing in a scenario where neutrinos are pseudo-Dirac \cite{Rink:2022nvw,Carloni:2022cqz}. Clearly point sources leave us with less statistics. However, the properties of the sources, e.g. the distance, flux and spectral index, are reasonably well known and we are not subject to uncertainties from the unknown source distribution.
In this, work we investigate the impact of secret neutrino interactions on the fluxes of NGC 1068 and TXS 0506+056 and show that these can be used to derive quite stringent limits.

This work is organized as follows. In Sec.~\ref{sec:optical_depth}  we introduce our model for secret neutrino interactions and present the relevant cross sections before computing the associated optical depth. We remain agnostic regarding the absolute mass of the neutrinos and discuss both the case where all relic neutrinos are non-relativistic today and the case where the lightest neutrino remains relativistic until the present. The impact of the interactions on the neutrino flux measured by the neutrino telescopes is then the topic of Sec.~\ref{sec:fluxes}. We use these insights to estimate limits on the parameter space from the observation of neutrino emission from NGC 1068 and TXS 0506+056 in Sec.~\ref{sec:limits}. Finally, we summarize our results in  Sec.~\ref{sec:summary}. Some technical details regarding our computations are given in the Appendix.


\section{Secret neutrino interactions and the optical depth}
\label{sec:optical_depth}

In this work, we take the neutrinos to be Majorana fermions\footnote{As far as this work is concerned, the differnce between Dirac and Majorana neutrinos is minor. However, Dirac neutrinos with secret interactions are strongly constrained by BBN since the new interaction can lead to a population of the right-handed neutrino component prior to BBN. This conflicts with the upper limit on $N_{\textrm{eff}}$ unless further new physics is invoked.} and introduce a secret neutrino interaction that is mediated by a new light scalar $\phi$. The interaction Lagrangian reads
\begin{align}
    \mathcal{L}_{int}= \frac{1}{2} \sum_{i,j} y_{ij}\bar{\nu}_i \nu_j \phi\,,
\end{align}
where $i$ and $j$ are flavor indices and $y_{ij}$ is a matrix of coupling constants.
In the following we assume the secret interaction to be flavor universal for the sake of simplicity. In this case, the neutrino scattering cross section is given by 
\begin{equation}
    \sigma_{\nu\nu}(s) = \frac{y^4}{32 \pi ((m_\phi^2-s)^2+m_\phi^2 \Gamma_\phi^2)s^2}\left( \frac{s(5 m_\phi^6-9 m_\phi^4 s + 6 s^3)}{m_\phi^2+s}+ \frac{2(5 m_\phi^8-9m_\phi^6 s +4 m_\phi^2 s^3)\log(\frac{m_\phi^2}{m_\phi^2+s})}{2 m_\phi^2+s}\right)\,,
    \label{eq:TreeLevelCrossSection}
\end{equation}
where $\Gamma_\phi= 3 y^2 m_\phi/(16 \pi)$ is the width of the scalar mediator and $s$ the usual Mandelstam variable. 
If $E_{\textrm{CM}}\geq m_\phi$ the production of a $\phi$-pair becomes possible. The cross section reads
\begin{align}
    \sigma_{\phi\phi}(s)=\frac{y^4}{64 \pi s^2}\left(\frac{s^2-4m_\phi^2 s +6 m_\phi^4}{s-2 m_\phi^2} \log \left[ \left( \frac{(s(s-4m_\phi^2))^{1/2}+s-2m_\phi^2}{(s(s-4m_\phi^2))^{1/2}-s+2m_\phi^2} \right)^2 \right]  -6 (s(s-4m_\phi^2))^{1/2}  \right)\,.
    \label{eq:phiphi_pair}
\end{align}
Our results for both processes agree with those reported by \cite{Esteban:2021tub} up to a factor of 2 which is due to a convention and drops out in the rate. We have also compared the matrix elements with those of \cite{Bringmann:2022aim} where a similar model is considered in a different context and find agreement. We cannot reproduce the neutrino scattering cross section of~\cite{Kelly:2018tyg}.
This interaction and the associated cross section should only be understood as an example and we expect other coupling structures or a different spin of the mediator to lead to qualitatively similar conclusions.

The astrophysical neutrinos can interact with the \CNB.
These neutrinos are a relic from the early universe. They decoupled from the rest of the SM at $T\sim 1\,\textrm{MeV}$, which is roughly the temperature when the interaction rate of neutrinos with the SM fell below the Hubble rate.
Today the \CNB has a thermal distribution with $T_{\nu}=1.9\,\textrm{K}$ 
and a total number density $n_{\textrm{tot}}\approx 340 \,\mbox{cm}^{-3}$.

\begin{figure}[t]
    \centering 
    \includegraphics[width=0.45\textwidth]{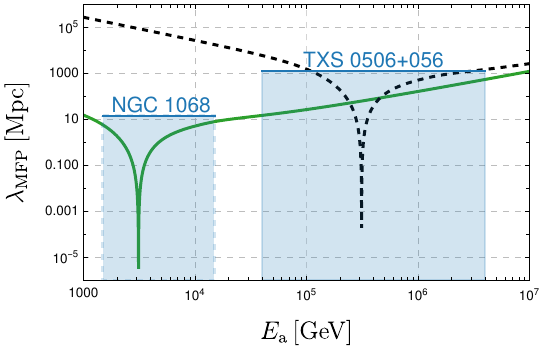}
    \includegraphics[width=0.45\textwidth]{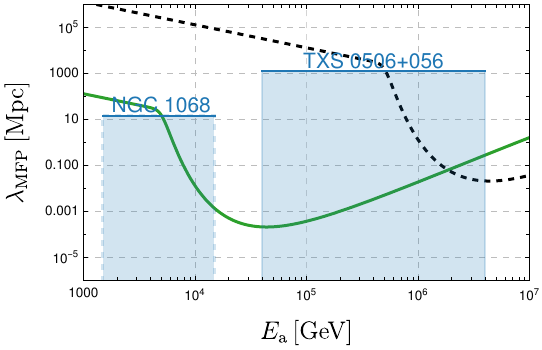}
    \caption{\textbf{Left:} The mean free path of an astrophysical neutrino as a function of its energy for two different mediator masses, $m_{\phi}=0.25\,\textrm{MeV}$ (green) and $m_{\phi}=2.5\,\textrm{MeV}$ (black, dashed). For illustration we assume a single neutrino species with mass $m_{\nu}=0.01\,\textrm{eV}$ and set $y=0.05$.  The distances of the two neutrino sources is indicated in blue for comparison. Redshift effects are neglected here.
    \textbf{Right:} Same as in the left panel but for $m_{\nu}\ll 1.9 \,\mbox{K} \sim 1.6 \times 10^{-4} \,\textrm{eV}$.
    }
    \label{fig:MFP}
\end{figure}

The key quantity to understand the effect on the flux is the transmittance $T$, i.e. the ratio of the received and the emitted flux. For a local source, such as NGC 1068, redshift can be neglected and the expressions are particularly simple. Then $T=e^{-\tau}$ where the optical depth $\tau$ of a source with distance $d$  can be determine via the mean free path $\lambda_{\textrm{MFP}}$  as $\tau=d/\lambda_{\textrm{MFP}}$. For ultra-relativistic particles the mean free path is simply given by the inverse of the interaction rate, $\Gamma=\lambda_{\textrm{MFP}}^{-1}$.
The interaction rate of the incident neutrino with energy $E_{\textrm{a}}$ with one mass eigenstate \CNB neutrino is given by
\begin{align}
\label{eq:fullrate}
 \Gamma_i(E_a)=\int \frac{\mathrm{d^3}p}{(2\pi)^3} f_i(\vec{p})\, v_{Møl}\sigma(s(E_{\textrm{a}},\vec{p}))\,,
\end{align}
where $f_i(\vec{p})$ is the momentum distribution of the \CNB neutrinos,  $v_{Møl}$ is the Møller velocity and $\sigma(s)=\sigma_{\nu\nu}(s)+\sigma_{\phi\phi}(s)$, see Appx.~\ref{Appx:massless_rate} for more details. 
A background neutrino with $m_i\gg T_\nu\approx 1.9\,\textrm{K}$ is non-relativistic today. In this case the center of mass energy is essentially independent of $\vec{p}$ with $s_{\textrm{nr}}\approx 2 E_{\textrm{a}} m_i$, $v_{Møl}=1$ and the  integral is trivially solved in the lab frame. Thus the rate for scattering on a population of non-relativistic background neutrinos reduces to
\begin{align}
 \Gamma_i(E_{\textrm{a}})= \sigma(2 E_{\textrm{a}}  m_i) n_i\,,
 \label{eq:Massive_Nu_Rate}
\end{align}
and one finds a simple expression for the optical depth
\begin{align}
   \tau(E_{\textrm{a}})= d \, \sum_{i=1}^3 n_i\, \sigma_{}( 2 E_{\textrm{a}}  m_i)\,.
\end{align}
For illustration, we show the mean free path in a simplified case for a single massive neutrino species and two different mediator masses in the left panel of Fig.\,\ref{fig:MFP}.
As expected, $\lambda_{\textrm{MFP}}$ inherits the resonance structure of $\sigma$ and we observe a strong dip when $m_\phi^2\approx 2 E_a m_i$.
The distances of the two known sources and the energy range of the observed emission is highlighted to put these numbers into perspective. 
 For the chosen benchmark values of the mediator mass of $0.25\,\textrm{MeV}$ and $2.5\,\textrm{MeV}$ with a coupling of $y=0.05$ and neutrino mass $m_i=0.01$~eV, we find that the mean free path falls below the distance to the two sources in the energy ranges $\sim 10^3 \; \mbox{to}\; 10^4\,\textrm{GeV}$ and $\sim 10^5 \; \mbox{to}\; 10^6\,\textrm{GeV}$ , respectively. Therefore, we expect a significant reduction of the flux of these sources.

The observed mass-squared splitting of the neutrinos \cite {Esteban:2020cvm} ensure that at least two mass eigenstates are non-relativistic today. However, the absolute mass scale of the neutrinos is not know and, therefore, one neutrino mass eigenstate could still be relativistic today. In this case, the full rate given in Eq.~\ref{eq:fullrate} should be used instead. We discuss how this rate can be computed efficiently and present useful approximate expressions in Appx.~\ref{Appx:massless_rate}. The right panel of Fig.\ref{fig:MFP} shows an illustration of the effect. We  use the same benchmark values as in the left panel except for the neutrino mass which is taken to be negligible compared to the temperature. We observe that the minimum of the mean free path is shifted to higher energies $E_a$ compared to the massive neutrino and also significantly broadened. 
This is simply due the fact that the typical center of mass energy $E_{\textrm{CM}}\sim\sqrt{T E_a}$ is smaller compared to the massive scenario and the wider range of center of mass energies that can occur in the collision of our test neutrino with a relativistic gas.  
In our example, the energy range where $\lambda_{\textrm{MFP}}$ is smaller than the distance extends over several orders of magnitude.
As can be guessed from theses figures, our ignorance of the absolute mass scale will have a strong impact on the interpretation of the astrophysical neutrino signal in our scenario.

For very distant sources this approach needs to be refined and the cosmological expansion has to be taken into account.  This can be done by solving a transport equation, see Appx.~\ref{Appx:transport_eq}. 
This yields an averaged optical depth \footnote{The shift of the flux due to redshift alone has been factored out here since it does not induce an observable effect at Earth.}
\begin{align}
    \tau(E_{\textrm{a}})=\int_0^z\frac{\Gamma(E_{\textrm{a}},z')}{(1+z^{\prime})H(z^{\prime})}\,\mathrm{d}z^{\prime},
    \label{eq:averaged_optical_depth}
\end{align}
where $z$ is the redshift of the source, $H$ is the Hubble rate as a function of redshift and $\Gamma$ denotes the total interaction rate of a neutrino with energy $E_{\textrm{a}}$ today at redshift $z'$. The $z'$-dependent rate can be obtained from the local expressions with the replacements $E_a \rightarrow E_a (1+z')$, $n_i\rightarrow n_i (1+z')^3$ and $T_\nu\rightarrow T_\nu (1+z')$.

\section{Impact on fluxes}
\label{sec:fluxes}
The effect of the new interaction depends crucially on the masses of the cosmic background neutrinos. 
Combining the results of various oscillation experiments, the mass square differences between the three SM neutrino mass eigenstates have been determined rather precisely. For normal ordering they are $\Delta m_{12}^2\approx0.75\cdot10^{-4}\,\textrm{eV}$ and $\Delta m_{23}^2\approx2.5\cdot10^{-3}\,\textrm{eV}$ \cite{Esteban:2020cvm}. 
However, the absolute scale of the neutrino masses remains unknown and only upper limits can be placed with lab experiments~\cite{KATRIN:2021uub}. Slightly more stringent bound comes from cosmology and sets a limit on the total sum of neutrino masses of $\sum_i m_i \le 0.12\,\textrm{eV}$~\cite{Planck:2018vyg}. This leaves us with the interesting possibility that the lightest neutrino is still relativistic today. Since a vanishing mass for the lightest eigenstate is predicted by some neutrino mass models, e.g. the minimal type I seesaw with two right-handed neutrinos \cite{Kleppe:1995zz,Ma:1998zg,Davidson:2006tg}, we include this possibility in our analysis.
In order to bracket our ignorance regarding the absolute mass scale of neutrinos we will therefore assume two benchmark scenarios:
\begin{itemize}
    \item BM1: $\sum_i m_i=0.1$ eV
    \item BM2: $m_1\ll   1.6 \times 10^{-4} \,\textrm{eV} \sim 1.9 \,\mbox{K} $
\end{itemize}
In both cases, we assume normal ordering (NO) and use the best-fit values for the  mass-squared splitting to determine the masses of the three eigenstates. The value chosen for BM1 is close  to the cosmological upper limit but does not saturate it. BM2 features one essentially massless neutrino eigenstate which leads to a strong broadening of the absorption feature as seen in the previous chapter.

\begin{figure}[t]
    \centering
    \includegraphics[width=0.30\textwidth]{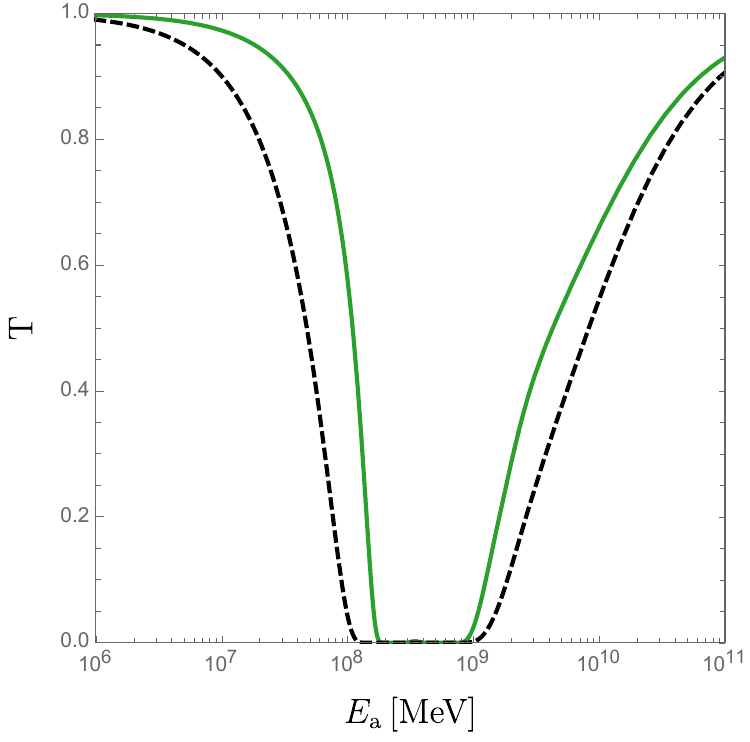} \qquad \includegraphics[width=0.30\textwidth]{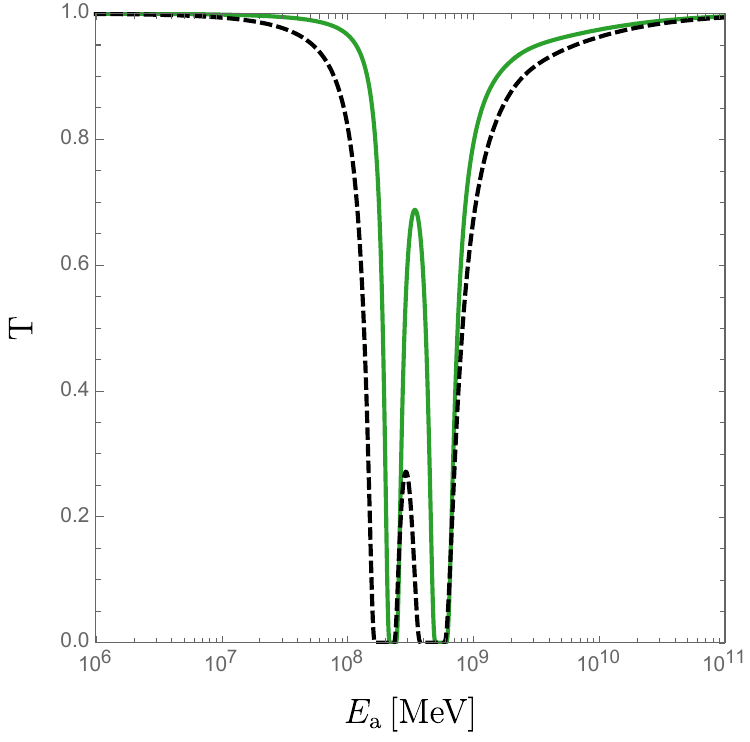}\qquad \includegraphics[width=0.30\textwidth]{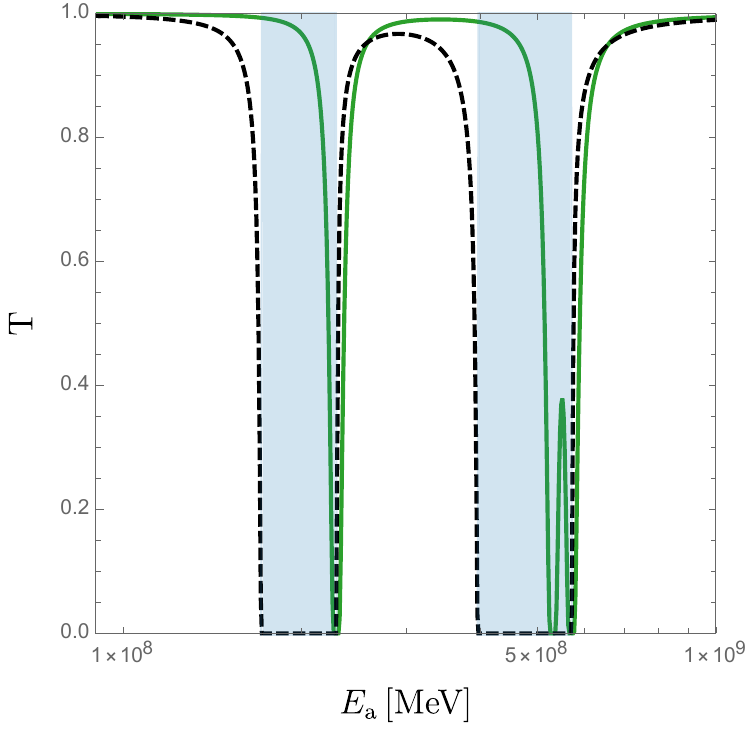}
    \caption{Transmittance $T$ as a function of $E_a$ for BM1, $m_\phi=5$ MeV and $d=1.2$ Gpc. The left, middle and right panel have $y=0.05$, $y=0.025$ and $y=0.01$, respectively. The solid green line shows the local result while the black dashed line takes redshift into account.  Note the different plot range in the right panel. Here the analytic estimate of the suppressed energy range is highlighted by the blue shading.}
    \label{fig:transmittance}
\end{figure}

Before discussing the flux, we take a look at the general features of the transmittance first. Since the results for a distant source without redshift correction are qualitatively similar to those of a local source we only discuss one exemplary source at $d=$ 1.2 Gpc, similar to TXS 0506+056.  
As can be seen in Fig.~\ref{fig:transmittance} its features depend strongly on the coupling. 
For sufficiently large values of the coupling, the adsorption due to the different neutrino mass eigenstates overlaps and we find one, sizable adsorption feature that leads to a strong suppression of the flux over more than one order of magnitude in $E_a$. 
For smaller values the range in which neutrinos are absorbed shrinks and the contributions from the individual mass eigenstates can be distinguished. In both cases, redshift leads to a broadening of the energy range in which the flux is suppressed.  
For even smaller $y$ only the most strongly resonance enhanced part of the cross section lead to an appreciable optical depth and we find  line-like absorption features at $E_a=m_\phi^2/2m_i$. 
Here the effect of redshift is very simple: the line gets broadened to the range
$m_\phi^2/2 m_i(1+z_{source}) \leq E_a\leq m_\phi^2/2 m_i$ since every neutrino that hits the resonance somewhere on its path from the source to Earth gets absorbed. 
In this regime, the signal is roughly independent of $y$. To spin this speculation even further, we can ask what the smallest coupling is that leads to a non-vanishing effect on the spectrum. 
One can determine this by using the narrow width approximation for the cross section and demanding the optical depth computed with Eq.~\ref{eq:averaged_optical_depth} is less than one \footnote{A simpler but similar estimate can be found by demanding that the peak cross section $\sigma_{max}\sim 1/m_\phi \Gamma$ is comparable to the target area $(n_i\cdot l_H)^{-1}$ where $l_H=1/H_0$ is the Hubble length.}.
This leads to a lower limit of $y\gtrsim 2\times 10^{-5} m_\phi[\mbox{MeV}]$ on the range of couplings that can hypothetically be tested using an instrument with perfect energy resolution and high statistics. 
Clearly, the present at IceCube is very far from this highly idealized situation and, therefore, we abandon this line of thought here.  For BM2, the energy range where the flux is significantly reduced is generically much wider and can easily extend over several orders of magnitude in energy. Therefore, this scenario frequently leads to an almost complete extinction of a source.

We now want to focus on more realistic situations and study the effect on the fluxes of the two known neutrino point sources, NGC 1068 and TXS 0506+056, as examples. We follow the IceCube collaboration and describe the unattenuated flux as 
\begin{align}
    \Phi_0(E)=\hat{\Phi}_0\cdot\left(\frac{E}{1\,\textrm{TeV}}\right)^{-\gamma}\,,
\end{align}
where $\gamma$ is the spectral index and $\hat{\Phi}_0$ parameterizes the intensity. 
NGC 1068 is a relatively nearby source with a distance~$d$ of $14.4$ Mpc. Therefore, the expansion of the Universe does not have an appreciable impact on it. The source spectrum is soft, with a best fit $\gamma=3.2$, and a normalization $ \hat{\Phi}_0= 5\times 10^{-11} \,\mbox{TeV}^{-1} \mbox{cm}^{-2} \mbox{s}^{-1}$. The emission is mainly observed in the energy window $1.5$ to $15$ TeV and is consistent with a constant source. TXS 0506+056 is much more distant and its redshift is measured to be $z=0.336$ \cite{Paiano:2018qeq}. This translates to a proper distance of $\sim1.2$ Gpc assuming a standard cosmology. 
The time-integrated search reports a best fit spectral index $\gamma=2.0$~\footnote{TXS 0506+056 is a transient source and an analysis that takes the time correlation with gamma-ray emission into account prefers a slightly higher value of $\gamma=2.2$ \cite{IceCube:2018dnn}. Nevertheless, we stick with the values preferred by \cite{IceCube:2022der} here.} and $\hat{\Phi}_0 =1.2 \times 10^{-13}\,\mbox{TeV}^{-1} \mbox{cm}^{-2} \mbox{s}^{-1}$. The main evidence for the emission comes from the energy range $40$ to $4000$ TeV. We assume these values for the sources and restrict our analysis to their respective energy ranges.

\begin{figure}[t]
    \centering 
    \includegraphics[width=0.49\textwidth]{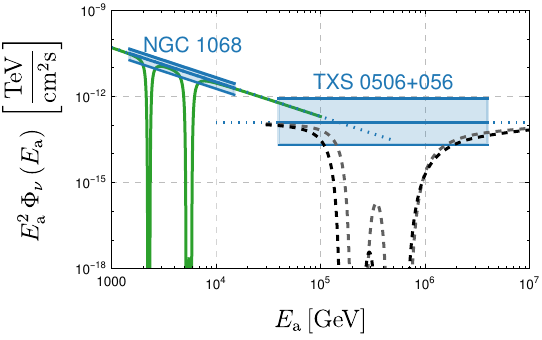}
    \includegraphics[width=0.49\textwidth]{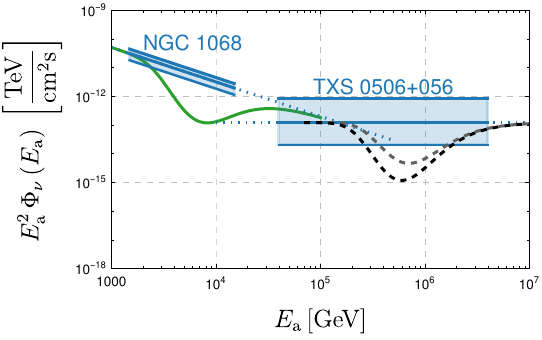}
    \caption{Flux of astrophysical neutrinos as a function of their energy, $E_{\textrm
    a}$. The blue solid lines illustrate the measured fluxes from AGN NGC 1068 and the blazar TXS 0506+056 by the IceCube collaboration together with their error margin.
    With the green-solid lines and black-dashed lines we show the theoretical expected fluxes from NGC 1068 and TXS 0506+056, respectively, for the presented model with different parameter choices. \textbf{Left:} 
    BM1, the coupling and the mediator mass are chosen to $m_{\phi}=0.5\,\textrm{MeV}$ $y=0.03$ for the solid-green curve and $m_{\phi}=5\,\textrm{MeV}$ $y=0.05$ for the black and gray-dashed curves. The gray-dashed curve is without redshift effect while it is included in the black one. \textbf{Right:} 
BM2, the model parameters are chosen to $m_{\phi}=0.1\,\textrm{MeV}$ and $y=2.5\cdot10^{-4}$ for the green curve and $m_{\phi}=1\,\textrm{MeV}$ and $y=2.5\cdot 10^{-4}$ for the black curve.}
    \label{fig:Flux}
\end{figure}

In Fig.\,\ref{fig:Flux} we show the expected neutrino flux from NGC 1068 for the first benchmark scenario with $m_{\phi}=0.5\,\textrm{MeV}$ and $y=0.03$  as green curve. Likewise, the predicted flux from the blazar TXS 0506+056 for the first scenario with $m_{\phi}=5\,\textrm{MeV}$ and $y=0.05$ is shown as black dashed line. In blue we present the measured fluxes of NGC 1068 and TXS 0506+056. Very prominent are the two absorption features from the three neutrino masses in this scenario. For the blazar, we indicate the effect of redshift by including the uncorrected result as a gray dashed line. Since the $z$ of TXS 0506+056 is not too big, the redshift only leads to a modest broadening of the suppressed energy range. 
For the second benchmark scenario, we find that the flux is strongly  reduced over a large range of energies. The expansion of the Universe also leads to a more extended absorption feature here, however, the impact is less pronounced than for BM1.

\section{Limits on secret interaction}
\label{sec:limits}
\subsection{Established sources}
\begin{figure}
    \centering
    \includegraphics{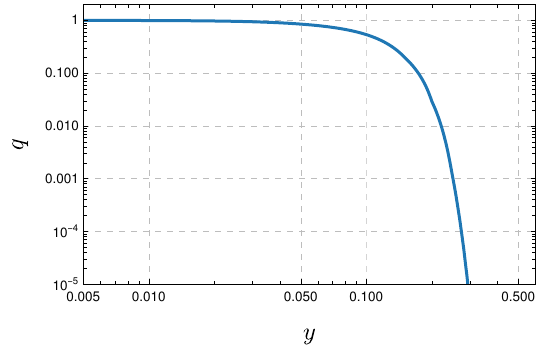}
    \caption{Dependence of $q$ on $y$ for a representative choice of $m_\phi=1$ MeV for NGC 1068. $\Phi_0(E)$ is taken to be the best fit spectrum reported by IceCube.}
    \label{fig:qofy}
\end{figure}

We estimate limits on the parameter space of our model by using the two sources, NGC 1068 and TXS 0506+056. At present, only a handful of neutrinos are associated with each of them. Therefore, a precise reconstruction of the  energy spectrum is not possible at the moment and the presence of spectral features is hard to constrain. 
Naturally, this limits our ability to constrain new physics with these observations.
Therefore, we focus on the total number of events here.
In order to relate the flux to the number of events in IceCube the effective area of the detector has to be taken into account.  The number of events in a given energy range $E_{\textrm{min}}$ to $E_{\textrm{max}}$ is given by 
\begin{align}
    n= t \int_{E_{\textrm{min}}}^{E_{\textrm{max}}} \textrm{d} E \, A_{\textrm{eff}}(E)\, \Phi(E)\,,
\end{align}
where $t$ is the exposure time and  $A_{\textrm{eff}}(E)$ is the effective area of the IceCube detector. For the position of TXS 0506+056 and NGC 1068 tabulated values of $A_{\textrm{eff}}(E)$ are included in the respective data releases \cite{TXS_data,NGC_data}. 
In order to derive a limit on the mediator mass and the coupling from these measurements we demand that the true number of neutrinos and the measured number of neutrinos should not differ by a value $q$
\begin{align}
    \frac{n}{n_0}=\frac{\int_{E_{\textrm{min}}}^{E_{\textrm{max}}} \textrm{d} E \, A_{\textrm{eff}}(E)\, \Phi(E)}{\int_{E_{\textrm{min}}}^{E_{\textrm{max}}} \textrm{d} E \, A_{\textrm{eff}}(E)\, \Phi_0(E)}\geq q\,.
\end{align}
In the following, we take $q$ to be $0.5$. This value is arbitrary and the results should only be interpreted as an estimate of the limit and not as an exclusion at some statistical significance. Nevertheless, we believe that this procedure is reasonable since the transmittance quickly transitions from line-like features to a strong suppression over a large energy range. This is illustrated in Fig.~\ref{fig:qofy} where we show the evolution of $q$ as a function of $y$ for a representative $m_\phi$. Once $q$ differs from 1 by an $\mathcal{O}(1)$ factor it starts to drop quickly with increasing $y$\footnote{This is also accompanied by $\mathcal{O}(1)$ changes in the spectral shape which makes the attenuated signal distinct from a simple power law.}. To compensate for this strong decrease of $q$ the true brightness of the source would need to increase by orders of magnitude if the observed number of events is to be realized in the presence of self-interactions. 

Naturally, the above argument could be strengthened significantly if a robust upper limit on the neutrino luminosity of the sources were available. At the moment, this  out of reach since the current understanding of the production mechanism and environment in both sources is still limited. While a naive comparison of the gamma-ray and neutrino luminosities of TXS 0506+056 is roughly consistent with the naive expectation of equal emission rates a  more sophisticated multi-messenger analysis of the photon spectrum encounter difficulties unless the production mechanism of the gamma-rays is partially disentangled from the neutrino one. Thus, the correlation between the high energy gamma-ray  and the neutrino luminosity is not a clear as expected.
In addition, \cite{Padovani:2019xcv} finds indications that TXS 0506+056 is not a blazar of the BL Lac type and suggests that there could be significant attenuation of the high energy photons. The situation is similar in the case of NGC 1068 in the the sense that a clear picture of the emission mechanism and the structure of the source is still being developed. Nevertheless, there are already first studies that indicate that accommodating an even higher  neutrino luminosity might be problematic. For example, \cite{Blanco:2023dfp} investigates how the neutrino luminosity and the soft gamma-ray spectrum below 1 GeV can be explained. They find that both a high proton luminosity and a high magnetic field are required to explain both observations. An even higher neutrino luminosity can thus only be accommodated with difficulty in their model.  In the future the cascade of the obscured high energy photons  down to MeV energies could be detected with novel instruments targeting the MeV/sub GeV photon spectrum, see e.g. \cite{Murase:2022dog}. Therefore, we expect that further observations and more careful modelling of the multimessenger signal of both sources will allow a more robust prior on the neutrino luminosity that can inform future searches for new physics in the neutrino sector. However, as the modelling of the emission from both sources is still far from settled and  we refrain from using strong assumptions about the relation between the photons and the neutrino luminosities.

\begin{figure}[t]
    \centering 
    \includegraphics[width=0.45\textwidth]{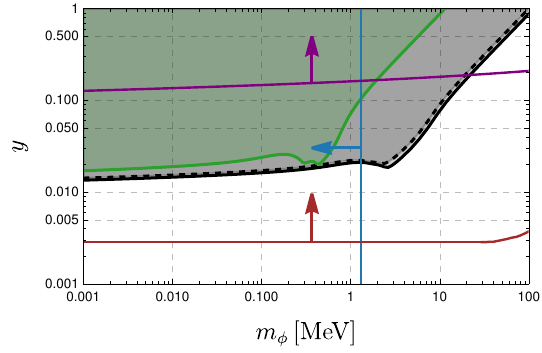}
    \includegraphics[width=0.45\textwidth]{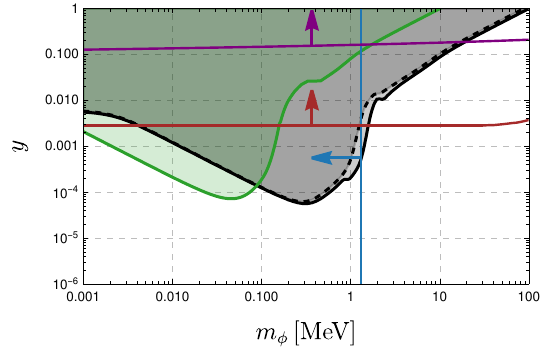}
    \caption{Limits on the secret coupling $y$ and the mediator mass $m_{\phi}$. The green and gray areas show the excluded regions inferred from the observation of the neutrino flux from NGC 1068 and TXS 0506+056, respectively. The black dashed line shows the constraint from TXS 0506+056 without redhsift effect. For comparison, the area constrained by BBN \cite{Blinov:2019gcj} is shown by the blue line attached with an arrow that points into the constrained direction. The brown and purple lines come from laboratory limits on $K^-$ \cite{Berryman:2018ogk} and $Z$ decay \cite{Brdar:2020nbj}.
    On the \textbf{Left} BM1 and on the \textbf{Right} for BM2. Note the different range of the y-axes.}
    \label{fig:Limits}
\end{figure}

With the criterion introduced above, we estimate the limits on the coupling $y$ as a function of the mediator mass $m_{\phi}$ for our two neutrino mass benchmarks. 
The bounds are depicted in Fig.\,\ref{fig:Limits}.
The left panel shows the limits for BM1. Here the green area shows the limit from NGC 1068 and the gray area the limit derived from TXS 0506+056. 
As can be seen, TXS 0506+056 is almost always more constraining than NGC 1068.  At low masses the advantage is only $O(1)$ since the lower energy of the neutrinos emitted by NGC 1068 partially compensates for the larger distance of TXS 0506+056 in the computation of the optical depth. The limits improve in both cases when the resonance condition $2 E_a m_i= m_\phi^2$ can be fulfilled in the energy range where the astrophysical neutrinos are observed. This allows the NGC 1068 limit to briefly dip down to the one from TXS 0506+056 at $m_\phi \approx0.5$ MeV. At high masses the TXS 0506+056 limit is clearly superior since both the larger distance and the higher energies enhance it compared to NGC 1068. However, the bound quickly deteriorates in this regime and for $m_\phi \gtrsim 100$ MeV only $y\gtrsim 1$ can be tested.
The right panel shows the results for BM2. 
As can be seen, the limits arising from the latter scenario are much more stringent.  This is due to the much larger range of neutrino energies $E_{\textrm{a}}$ that allow $\phi$  to go on-shell in the scattering process when the background is a relativistic gas.
In this scenario, couplings of $\sim  10^{-3}$ are excluded for mediator masses of $\sim 10^{-3}$ MeV.  For $m_\phi \gtrsim 1\,\textrm{MeV}$ the limits start to weaken strongly from $m_\phi\approx 60$ MeV on only $y\gtrsim 1$ can be excluded. In the intermediate region with mediator masses from $\mathcal{O}(1)$ keV up to a few MeV the limits are quite strong and go down to almost $5 \times 10^{-5}$. Here, NGC 1068 can improve considerably on TXS 0506+056  for masses below $0.1$ MeV.

Finally, it is interesting to compare these results with bound from lab experiments. For comparison, we have therefore included limits from $K\rightarrow{\mu \bar{\nu} \phi}$ decays following \cite{Berryman:2018ogk} and bounds from $Z$ decays \cite{Brdar:2020nbj} to Fig.~\ref{fig:Limits}. In this context it is important to note that the lab constraints depend strongly on the flavor structure of the interaction. The lab bounds are quite stringent for new physics coupling to $\nu_\mu$ but become rather weak in the case of a coupling to $\nu_\tau$. As can be seen the astrophysical limit are comparable with the lab bound on new interactions of $\nu_\mu$ in the region $m_\phi \approx 1$ MeV for an approximately massless lightest neutrino and lag considerably behind elsewhere. The situation is different for new physics that couple preferentially to $\nu_\tau$ since in this case the lab bound is only the ballpark of a few $10^{-1}$ and astrophysical limits can provide the most stringent limit. 

As the laboratory constraints are strongly flavor-dependent it is interesting to ask how our bounds would be affected if the couplings are not universal, as we have assumed so far, but rather flavor-specific. 
In particular, a tauphilic scalar field is an interesting possibility since laboratory bounds on this interaction lag considerably behind the others. Using some approximations, see Appendix\,\ref{Appx:tau_limits} for details, we have found two simple rules to translate the limits on a universal coupling $y_u$ into one on a tauphilic Yukawa $y_\tau$. In regions of parameter space where $\sigma\sim y^4$ dictates the reduction in the flux, we find $y_u/y_{\tau} \approx 0.58$. In regions of parameter space where the resonant peak dominates the attenuation, e.g. where scattering on a massless lightest neutrino is dominant, the ratio is $y_u/y_{\tau}\sim\sqrt{|U_{\tau 1}|^2/3}\approx 0.28$ instead. To check these estimates we have rerun our analysis with a tauphilic coupling structure and find excellent agreement, see Fig. \ref{fig:MN_TauFlavor} and Fig. \ref{fig:limit_NGC_massless_tauFlav}.

\subsection{Tentative signal from PKS 1424+240}
\label{sec:outlook}
\begin{figure}[t]
    \centering 
    \includegraphics[width=0.45\textwidth]{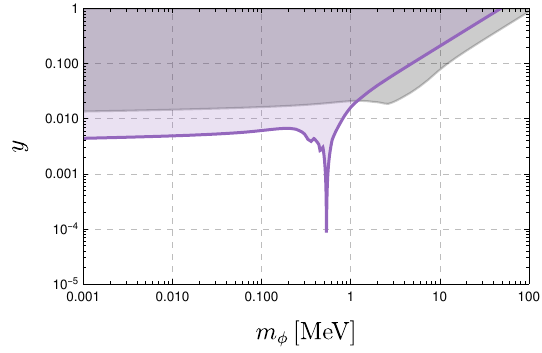}
    \includegraphics[width=0.45\textwidth]{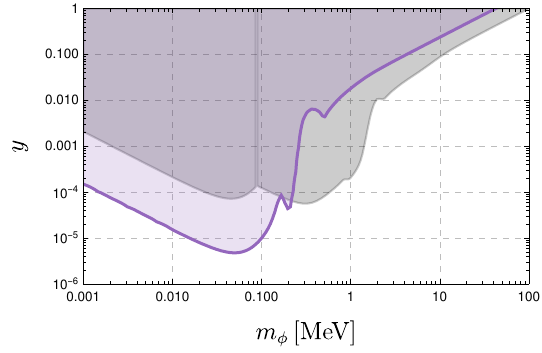}
    \caption{Estimated limits on the coupling $y$ vs. the scalar mass $m_{\phi}$ for PKS 1424+240 in our two benchmark models, BM1 on the \textbf{left} and BM2 on the \textbf{right}. The grey shaded areas are the combined estimates from NGC 1068 and TXS 0506+056 from Fig.\,\ref{fig:Limits}.}
    \label{fig:LimitsPKS}
\end{figure}
In the future, more sources are expected to be detected and join TXS 0506+056 and NGC 1068. It is interesting to speculate what impact they may have on the limits on secret neutrino interactions. Currently, the most promising candidate for an additional neutrino point source is PKS 1424+240. IceCube observed a neutrino signal associated with it with a local significance of $3.7\sigma$, see \cite{IceCube:2022der} and supplementary material thereof. While this does not qualify as a detection yet, it is well above all other source candidates considered in this search except NGC 1068 and TXS 0506 +056.  Interestingly, it is also a very far away source, and its redshift of $z=0.60$~\cite{Rovero:2016igo} or, equivalently, a proper distance of  $d\approx1.8\,\textrm{Gpc}$ even exceeds that of TXS 0506+056. Its best-fit spectral index is
$\gamma_{\textrm{PKS}}=3.5$  
which is rather similar to NGC 1068 such that the energy range in which the signal is detected is comparable. 

To assess how a robust establishment of this source will affect the limits we employ the analysis outlined above to PKS 1424+240. The results are shown in Fig. \ref{fig:LimitsPKS}.
There are some notable features. First off, as expected from the similar spectral index, the overall shape of the tentative limits have some similarity with those from NGC 1068. However, their overall strength is significantly enhanced, and the upper bound on $y$ decreases by about one order of magnitude. This brings the limits into the vicinity of the kaon limits for low $m_\phi$ and allows them to almost catch up to TXS 0506+056 at high masses.
A slightly surprising feature in Fig. \ref{fig:LimitsPKS} is the appearance of the very pronounced peak at $m_\phi\approx0.5$ MeV in the left panel, and, to a lesser degree the more pronounced downward spikes in the left half of the right panel. These can be understood due to the effect of redshift which is more important in this case than for the sources discussed previously. The spikes are caused by the redshift-broadening of the resonantly enhanced absorption throat discussed in Sec. II (see in particular Fig. 2 for an illustration of the effect). With a redshift of $z=0.6$ and a relatively soft spectrum, this is sufficient to strongly dampen away the signal in about half the relevant energy range for a relatively small range of masses that maximize the effect. Thus we can meet the requirement $q=0.5$ without relying on the off-shell part of the cross section. Clearly, the absence of the signal in such a large energy range would not be consistent with a power-law spectrum and could be detected even before the high-quality spectra become available that would enable a similar exercise for lower redshift sources.

\section{Summary and conclusions}
\label{sec:summary}
The recent discovery of a steady neutrino point source in NGC 1068 by the IceCube Collaboration opens a new chapter for neutrino astronomy which offers a large range of new opportunities. 
Naturally, the most obvious and most important is a new view of the Universe at the highest energies. This is expected to pave the way to a better understanding of the sources of cosmic rays. 
In addition, high energy neutrinos that have traversed well-measured astrophysical distances offer further opportunities and can be used to test physics beyond the SM.
In this work, we studied the implications of secret neutrino interactions on the signal from neutrinos point sources. 
We use a minimal parameterization of such an interaction based on a light new scalar mediator with a $\mathcal{O}(\mbox{MeV})$ mass that couples to neutrinos. 
This allows for interactions of the astrophysical neutrinos with the \CNB that scatter them out of the path to Earth and thus lead to an energy- and distance-dependent reduction of the observed flux. 
The key ingredient to analyze this phenomenon quantitatively is the ratio of the distance and the mean free path, i.e. the optical depth~$\tau$. 
It is controlled by the new physics parameters but it also depends on unknown SM physics: the absolute mass scale of the background neutrinos. 
In particular, if the lightest mass eigenstate is still relativistic today the optical depth can be significantly enhanced compared to the case where all species of the \CNB are non-relativistic. 
This is due to the fact that the condition for a resonant enhancement of the cross section can be met over a larger range of energies. 
We provide simple, analytic expressions for the optical depth of local sources in both cases and discuss the averaging procedure that is needed if a source is far enough away for the expansion of the Universe to play a role. 
Equipped with these general results we analyze the impact on the flux of the two know sources of astrophysical neutrinos, NGC 1068 and TXS 0506+056, and estimate limits on the parameter space of the model. 
For low mass mediators, the limits inferred from both sources are comparable while for high masses of an MeV or more TXS 0506+056, which has a harder spectrum and is observed at higher energies, is more constraining. In the intermediate region, the sensitivity is boosted if the resonant enhancement of the cross section falls in the energy range where neutrinos from a source are emitted. 
This effect is much more pronounced if the lightest neutrino is approximately massless and we find a significant improvement of the sensitivity compared to a non-relativistic \CNB. 
Thus a robust determination of the absolute mass scale is highly desirable for proper assessments of the potential to test new physics with point sources. 
However, with the discovery of additional sources and more precise determinations of the flux, we expect that significant improvements will be possible in the near future even if the absolute masses of the neutrinos should remain elusive. One highly promising candidate for this is PKS 1424+240. As this source is even further away than TXS 0506+056, it could strengthen the bounds considerably. With, more data  on the way the prospects for testing new physics with new astrophysical neutrino points source is exciting.

 \section*{Acknowledgements}
The work of CD is supported by the F.R.S./FNRS under the Excellence of Science (EoS) project No.\ 30820817 - be.h ``The H boson gateway to
physics beyond the Standard Model'' and by the IISN convention No.\ 4.4503.15.

\appendix

\section{Rate for massless neutrinos}
\label{Appx:massless_rate}

\begin{figure}[t]
    \centering 
    \includegraphics[width=0.6\textwidth]{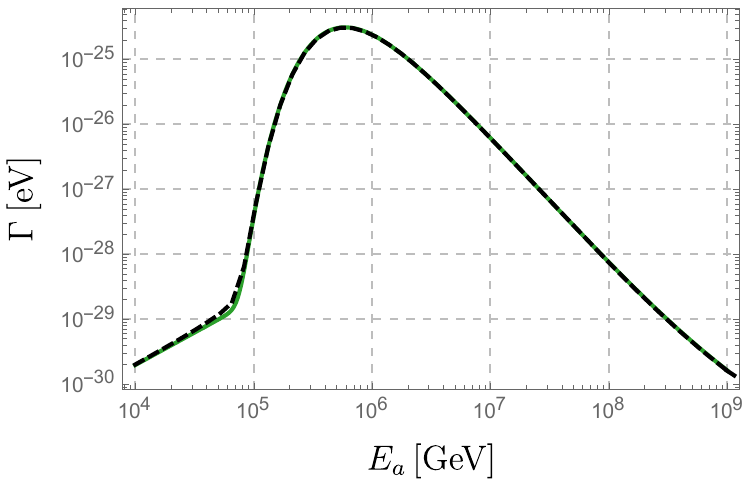}
    \caption{ Analytic approximation of $\Gamma_{\nu \nu \rightarrow\nu \nu}$  for a gas of massless background neutrinos as a function of incident neutrino energy $E_{\nu}$ (green, solid). The model paramters are set to $y=0.2$ and $M_{\phi}=1\,\textrm{MeV}$. For comparison, the  full numerical result is shown as the black dashed line.}
    \label{fig:Rates_AnavsNum}
\end{figure}

The rate for the interaction of a neutrino with energy $E_a$ with the \CNB neutrinos is given by
\begin{align}
\Gamma(E_a)= \int \frac{\mathrm{d}\vec{p_2}}{(2\pi)^3} f(|\vec{p}_2|)\, v_{Møl} \sigma_{\nu\nu}(E_{a},\vec{p_2})\,.
\end{align}
Here $f(|\vec{p}|)$ is the Fermi-Dirac distribution $f(|\vec{p}|)=1/(\exp[p/T_\nu]+1)$ and $\sigma_{\nu\nu}(E_{\nu},\vec{p_2})$ is the neutrino-neutrino cross section for an incident neutrino with energy $E_{a}$.
The Møller velocity is given by 
$v_{Møl}=F/(E_a E_2)$ where $F =\sqrt{(p_a \cdot p_2)^2-m_a^2 m_2^2}$ is the flux factor \cite{Gondolo:1990dk}.
This integral can be rewritten using the simplifications outlined in \cite{Ala-Mattinen:2019mpa}. For massless neutrinos one finds 
\begin{align}
 \Gamma(E_{\textrm{a}})=\frac{1}{16\pi^2 E_{\textrm{a}}^2}\int_0^{\infty}\mathrm{d}E_2\,\frac{g_{i}}{e^{\frac{E_2}{T_{\nu}}}+1}\int_{s_-=0}^{s_+=4E_2 E_{\textrm{a}}}\mathrm{d}s\, s\,\sigma_{\nu\nu}(s)\,,
\end{align}
where $g_i=2$ denotes the internal degrees of freedom of one Majorana neutrino species.
For convenience we report some approximate expressions for $\Gamma$ in the following. If $m_\phi \gg \sqrt{E_{\textrm{a}} T_\nu}$ 
the resonance does not play a role and the interaction can be modeled as a contact interaction. In this case we find
\begin{align}
    \Gamma_{\textrm{heavy}}\approx \frac{7 \pi^3 y^4 }{2592 \,\zeta(3)}\frac{ E_{\textrm{a}} T_\nu}{m_\phi^4}n_{\nu_1}\,,
\end{align}
 where $\zeta(3)$ is the Riemann $\zeta$ function evaluated at 3. In the opposite limit, for a negligible $m_\phi$ we find  
\begin{align}
    \Gamma_{\textrm{light}}\approx \frac{\pi y^4 }{192 \,\zeta(3) }  \frac{1}{E_{\textrm{a}}\, T_\nu}n_{\nu_1}\,.
\end{align}
The most interesting situation arises when $m_\phi$ and $\sqrt{E_\nu T_\nu}$ are of the same order. In this case the resonance condition can be met for some of the \CNB~neutrinos and one finds a large enhancement of the rate. Using the narrow width approximation (NWA) for the resonant contribution we get 
\begin{align}
    \Gamma_{\textrm{NWA}}\approx\frac{y^4 }{384 \zeta(3)  } \frac{m_\phi^3 }{E_\nu^2\, T_\nu^2 \,\Gamma_\phi} \log[1+e^{-\frac{m_\phi^2}{4 E_\nu T_\nu}}] n_{\nu_1}\,.
\end{align}
A reasonable approximation of the full rate can be found by switching from $\Gamma_\textrm{heavy}$ to $\Gamma_\textrm{light}$ at the point where they are equal to capture of the off-shell contribution and adding the on-shell contribution as described by $\Gamma_{\textrm{NWA}}$ on top. We show a comparison of this approach with the full numerical integral in Fig.\,\ref{fig:Rates_AnavsNum}. As can be seen the agreement is very good.

Finally, $\phi$ pair production can contribute to the rate. For $m_\phi^2\ll E_a T_\nu$ it can be approximated as
\begin{align}
    \Gamma_{\phi\phi,\textrm{light}}\approx \frac{y^4}{48 \pi \zeta(3) }\frac{\log\left[\frac{4 E_\nu T_\nu}{m_{\phi}^2} \right] -\gamma}{E_\nu T_\nu}n_{\nu_1}\,,
\end{align}
where $\gamma$ is the Euler–Mascheroni constant.

\section{Transport equation}
\label{Appx:transport_eq}
In an expanding universe the differential neutrino number density  evolves according to a transport equation \cite{Ng:2014pca,DiFranzo:2015qea,Esteban:2021tub}, see \cite{Bhattacharjee:1999mup} for a more general review in the context of general cosmic ray transport. Neglecting neutrino reproduction, which only plays a rule for the diffuse neutrino background, it reads
\begin{align}
 \frac{\partial\Phi (t,E_{\textrm{a}})}{\partial t}=\frac{\partial}{\partial E_{\textrm{a}}} [H(t) E_{\textrm{a}} \Phi(t,E_{\textrm{a}})] -\Phi(t,E_{\textrm{a}})\Gamma(E_{\textrm{a}},t)\,,
 \label{eq:NeutrinoTransport}
\end{align}
where $\Gamma(E_{\textrm{a}},t)$ should be understood as the interaction rate of a neutrino with energy $E_{\textrm{a}}$ today at time $t$. 
This equation can be simplified by rewriting the neutrino flux as
\begin{align}
 Z(z,E_{\textrm{a}}):=(1+z)\Phi(z,E_{\textrm{a}}[1+z])\,,
\end{align}
where $z$ is the redshift. With $\frac{\partial}{\partial t}=-H(z)\cdot (1+z)\frac{\partial}{\partial z}$ this
changes it to
\begin{align}
\label{eq:transport_simplified}
\frac{\partial Z(z,E_{\textrm{a}})}{\partial z}=\frac{Z(z,E_{\textrm{a}})\Gamma(E_{\textrm{a}},z)}{H(z)(1+z)}\,.
\end{align}
In the range of $z$ relevant for astrophysical neutrino sources $H(z)\approx H_0 (\Omega_\Lambda +\Omega_m (1+z)^3)^{1/2}$ where $H_0$ is the Hubble constant, while $\Omega_\Lambda$ ($\Omega_m$) is the ratio of the energy density of dark energy (matter) to the critical energy density in the Universe today. We use the best fit values reported by Planck \cite{Planck:2018vyg} in our numerical computations. 
The $z$-dependent rate is given by 
\begin{align}
 \Gamma_i(E_{\textrm{a}},z)=\int \frac{\mathrm{d^3}p}{(2\pi)^3} (1+z)^3 f_i(\vec{p}(1+z))\, v_{Møl} \sigma_{\nu\nu}(s(E_{\textrm{a}}(1+z),\vec{p}(1+z)))\,,
\end{align}
where $p$ and $E_a$ should be understood as the values today. For a background neutrino species that remains non-relativistic in the whole range of $z$ this reduces to $\Gamma_i(E_a,z)=\sigma(2 m_i E_a (1+z)) n_i (1+z)^3$, i.e. we correct for the redshift of the neutrino energy, $E_a \rightarrow E_a (1+z)$, and the \CNB density, $n_i \rightarrow{n_i (1+z)^3}$. For a massless neutrino one needs to account for the redshift of the neutrino temperature as well, i.e. $T_\nu \rightarrow T_\nu (1+z)$. With these replacements the results in Appx. \ref{Appx:massless_rate} can be used directly.

 Recasting Eq.~\ref{eq:transport_simplified} as an integral equation we find for the transmittance of a source at redshift z
\begin{align}
    T=\frac{Z(0,E_{\nu})}{Z(z,E_{\nu})}=\mathrm{Exp}\left[-\int_0^z\frac{1}{H(z^{\prime})(1+z^{\prime})}\Gamma(E_{\nu},z')\,\mathrm{d}z^{\prime}\right]\,.
    \label{eq:Solution_TE}
\end{align}
Interpreting the integral in the exponential as an averaged optical depth then leads to the expression in the main text.

\section{Converting limits for universal couplings to flavor specific ones}
\label{Appx:tau_limits}

\begin{figure}[t]
    \centering 
    \includegraphics[width=0.45\textwidth]{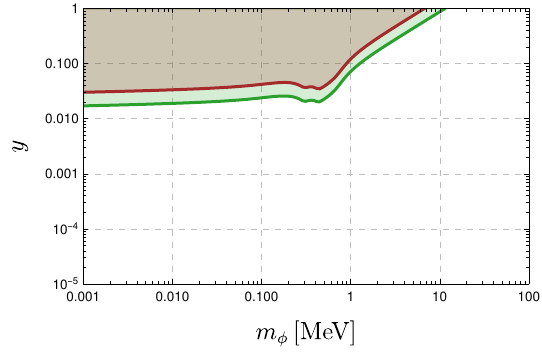}
    \includegraphics[width=0.45\textwidth]{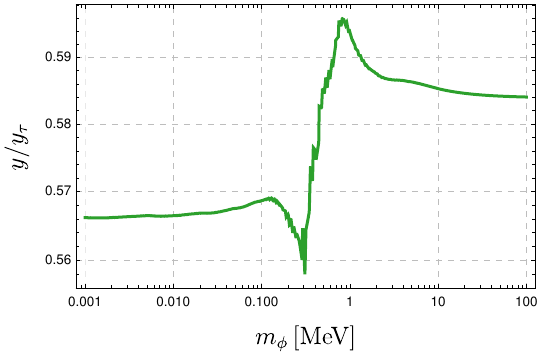}
    \caption{The \textbf{left} plot shows the estimated limit on the scalar coupling as a function of the scalar mass derived from NGC 1068. The green curve is the same as in Fig.~\ref{fig:Limits}, the dark red curve shows the limit derived for a flavor-depended coupling to $\nu_{\tau}$ only. The \textbf{right} plot shows the ratio of the universal coupling $y$ to the coupling $y_\tau$ for the tauphilic scenario.}
    \label{fig:MN_TauFlavor}
\end{figure}

\begin{figure}[t]
    \centering 
    \includegraphics[width=0.45\textwidth]{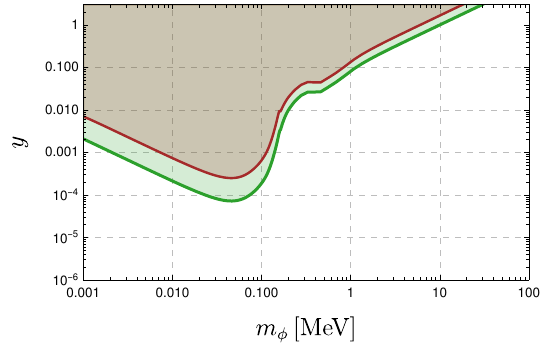}
    \includegraphics[width=0.45\textwidth]{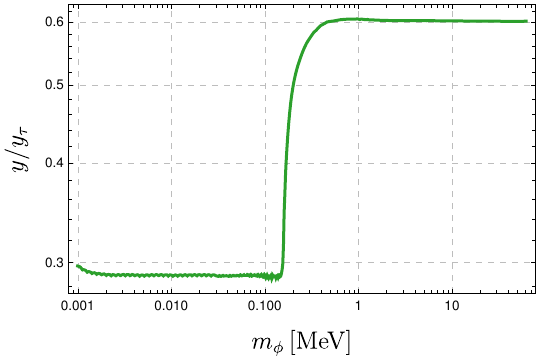}
    \caption{Same as Fig.~\ref{fig:MN_TauFlavor} for BM2.}
    \label{fig:limit_NGC_massless_tauFlav}
\end{figure}

In the case of $\tau$-specific couplings our limits would exclude new parts of parameter space. Therefore, it would be interesting to have a simple conversion formula at hand that allows to recast the universal limits as a flavor specific one. In order to achieve this one can compare the fluxes observed under the two different hypothesis for the coupling structure. In full generality, the observed flux is
    \begin{align}
        \phi_{obs}=   \sum_{i} \exp\left( -\tau_{i}\right) \frac{\phi_{source}}{3}\,,
    \end{align}
    where $i=1,2,3$ denotes the mass eigenstate of the incoming neutrino and we exploited that due to the large distance the neutrino eigenstates are decohered over the bulk of their propagation. In our scenario, i.e. for universal diagonal couplings, this reduces to 
    \begin{align}
        \phi_{obs}= \exp(-\tau_{u})\phi_{source} \,,
        \label{eq:convertlimits1}
    \end{align}
    where we have introduced the subscript $u$ to stress that the expressions have to be computed with a universal coupling. As in our main text $\tau_u= d \sum_j \sigma_u n_j$,
    and $j$ indicates the mass eigenstate of the target neutrino (the generalization to massless neutrinos and the inclusion of redshift is straightforward and does not add to this discussion so we omit these effects here).
    For a more general coupling structure we instead get
      \begin{align}
        \phi_{obs}=   \sum_{i} \exp\left( -d \sum_j \sigma_{ij} n_j\right) \frac{\phi_{source}}{3}\,.
        \label{eq:convertlimits2}
    \end{align}
    We would like to remind the reader that $\sigma_{ij}$ implicitly depends on $m_j$  here since $s=2E_i m_j$ for scattering on massive neutrinos.
    For a flavor specific coupling to $\tau$ neutrinos we have $\sigma_{ij}= |U_{i\tau}|^2 |U_{j,\tau}|^2 \sigma_\tau(s(m_j))$ 
    where $U_{ij}$ are the entries of the PMNS-matrix and the superscript on $\sigma_\tau$ indicates that the Yukawa $y=y_\tau$, i.e. that the flavor specific coupling is used. Equating the observed fluxes in Eq.\,(\ref{eq:convertlimits1}) and (\ref{eq:convertlimits2}) now allows to establish a relation between universal coupling $y_u$ and $y_\tau$.
    \begin{align}
        \exp\left(d\sum_j \sigma_u n_j\right)=\frac{1}{3}   \sum_{i} \exp\left( -d |U_{\tau i}|^2\sum_j |U_{j,\tau}|^2 \sigma_\tau n_j\right)\,.
    \end{align}
    Unfortunately, it is not possible to do so analytically in full generality. Nevertheless, one can go forward by using some simple approximations. Expanding the exponentials for small $\tau$ we find after exploiting the unitarity of the PMNS matrix
    \begin{align}
        \sum_j \sigma_u n_j= \frac{1}{3} \sum_j |U_{\tau j}|^2 \sigma_\tau n_j\,.
    \end{align}
    As the cross section is strongly energy dependent it is typically dominated by one target neutrino at a given energy such that one expects 
    $\sigma_u n_j\approx \frac{1}{3} |U_{j \tau}|^2 \sigma_\tau n_j$. 
    For $\sigma \propto y^4$ this 
    yields $\frac{y_u}{y_\tau}\approx \left( \frac{|U_{\tau j}|^2}{3}\right)^{1/4}$.     Using the best fit values for the mixing angles from \cite{deSalas:2020pgw} to compute the PMNS matrix elements, we find  $U_{\tau 1} = 0.49$ and $U_{\tau 2} = 0.57$ and $U_{\tau 3} = 0.67$. As the exclusion is set by averaging over an energy range one expects that the true conversion factor can be found by averaging over $j$ which just leads to $\frac{y_u}{y_\tau} \approx \sqrt{1/3}\approx 0.58$.
    To asses whether this estimate captures the relevant physics correctly we have rederived the limit in $y_\tau$ using the full expression outlined above for NGC 1068. The results are shown in Fig. \ref{fig:MN_TauFlavor}. As can be seen the estimate works very well and the conversion of the limits is to very good precision given by  $\frac{y_u}{y_\tau}\approx 0.58$.
Note, that this argument needs to be modified if the absorption is dominated by the resonantly enhanced part of the cross section as is the case for a massless neutrino. In this situation the scattering rate is proportional to $y^2$ and we thus expect  
$\frac{y_u}{y_\tau}\approx \sqrt{ \frac{|U_{\tau j}|^2}{3}}\approx 0.28$. As can be seen in Fig.~\ref{fig:limit_NGC_massless_tauFlav}, this describes the behavior of the limits very nicely in the case with a massless neutrino.
Similar rescaling laws for flavor specific coupling to $\nu_e$ and $\nu_{\mu}$ can be obtained completely analogously.

\bibliographystyle{utphys.bst}
\bibliography{bibliography.bib}

\end{document}